\documentclass[ApJL,twocolumn]{aastex62}
\usepackage{amssymb, amsmath, graphicx, dcolumn, color,units, xspace, mathtools, physics, tensor, bm, lipsum, revsymb}
\usepackage[normalem]{ulem}

\newcommand{\default}{\textsc{Default}\xspace}
\newcommand{\exdefault}{\textsc{Extended}\xspace}

\newcommand{\gaussian}{\textsc{Gaussian}\xspace}

\newcommand{\SPA}{School of Physics and Astronomy, Monash University, Clayton VIC 3800, Australia}
\newcommand{\OzGravMonash}{OzGrav: The ARC Centre of Excellence for Gravitational Wave Discovery, Clayton VIC 3800, Australia}

\shorttitle{Better black hole spin models}
\shortauthors{Galaudage \textit{et al}.}

\begin{document}

\title{Building better spin models for merging binary black holes: \\ Evidence for non-spinning and rapidly spinning nearly aligned sub-populations}

\author{Shanika Galaudage}
\affiliation{\SPA}
\affiliation{\OzGravMonash}
\correspondingauthor{Shanika Galaudage}
\email{shanika.galaudage@monash.edu}

\author{Colm Talbot}
\affiliation{LIGO Laboratory, California Institute of Technology, Pasadena, CA 91125, USA}

\author{Tushar Nagar}
\affiliation{\SPA}
\affiliation{\OzGravMonash}

\author{Deepnika Jain}
\affiliation{National Institute of Technology Karnataka, Surathkal, Srinivasnagar, 57025, India}

\author{Eric Thrane}
\affiliation{\SPA}
\affiliation{\OzGravMonash}

\author{Ilya Mandel}
\affiliation{\SPA}
\affiliation{\OzGravMonash}
\affiliation{Institute of Gravitational Wave Astronomy and School of Physics and Astronomy,\\
University of Birmingham, Birmingham, B15 2TT, United Kingdom}

\date{\today}

\begin{abstract}
Recent work paints a conflicting portrait of the distribution of black hole spins in merging binaries measured with gravitational waves.
Some analyses find that a significant fraction of merging binaries contain at least one black hole with a spin tilt $>90^\circ$ with respect to the orbital angular momentum vector, which has been interpreted as a signature for dynamical assembly.
Other analyses find the data are consistent with a bimodal population in which some binaries contain black holes with negligible spin while the rest contain black holes with spin vectors preferentially aligned with the orbital angular momentum vector.
In this work, we scrutinize models for the distribution of black hole spins to pinpoint possible failure modes in which the model yields a faulty conclusion.
We reanalyze data from the second LIGO--Virgo gravitational-wave transient catalog (GWTC-2) using a revised spin model, which allows for a sub-population of black holes with negligible spins.
In agreement with recent results by Roulet et al.,  we show that the GWTC-2 detections are consistent with two distinct sub-populations.
We estimate that {$29-75\%$} (90\% credible interval) of merging binaries contain black holes with negligible spin $\chi \approx 0$. 
The remaining binaries are part of a second sub-population in which the spin vectors are preferentially (but not exactly) aligned to the orbital angular momentum.
The black holes in this second sub-population are characterized by spins of {$\chi \sim 0.45$}.
We suggest that the inferred spin distribution is consistent with the hypothesis that all merging binaries form via the field formation scenario.
\end{abstract}

\section{Introduction}\label{sec:intro}
Gravitational waves from merging binaries encode information about the mass and spin of the component black holes and/or neutron stars.
These properties, in turn, provides clues as to how the binary formed.
Two scenarios are frequently invoked to explain merging binary black holes: field and dynamical (see \citealt{MandelFarmer:2018,Mapelli:2021, MandelBroekgaarden:2021} for reviews).
In the field scenario, binary black hole (BBH) systems are formed from isolated stellar binaries.  
In the dynamical scenario, BBH systems are assembled through interactions in dense stellar environments such as globular clusters and nuclear clusters; see \cite{Mapelli:2020} for a recent review.

These two scenarios yield distinct predictions for the distribution of black hole spin vectors.
Field binaries are generally expected to form BBH systems with dimensionless spin vectors $\vec\chi_1, \vec\chi_2$ preferentially-aligned with the orbital angular momentum vector $\vec{L}$.
Supernova kicks may serve to somewhat misalign the $\vec\chi_{1,2}$ and $\vec{L}$ vectors, but the typical misalignment angle is expected to be modest~\citep{OShaughnessy2017,Stevenson2017,Gerosa2018,Rodriguez2016,Bavera2020}.
Dynamically assembled binaries on the other hand are expected to form BBH systems with isotropically distributed spin vectors~\citep{Kalogera2000, MandelOShaughnessy2010, Rodriguez2016, Zevin2017, spin, Rodriguez2018, Doctor2019}.
These two predictions can be used to estimate the fraction of merging binaries associated with each channel.

Using data from the second LIGO--Virgo gravitational-wave transient catalog (GWTC-2) \citep{GWTC-2}, \cite{o3a_pop} fit the distribution of $\vec\chi_1,\vec\chi_2$ with two different models, which produced qualitatively similar results.
The \default model incorporates the spin magnitude model from \cite{Wysocki2019}:
\begin{align}\label{eq:chi}
    \pi(\chi_{1,2} | \alpha_\chi, \beta_\chi) = \text{Beta}(\chi_{1,2} | \alpha_\chi, \beta_\chi) .
\end{align}
Here, $\chi_{1,2}\equiv|\vec\chi_{1,2}|$ are the magnitudes of the dimensionless spin vectors, $\pi(...)$ denotes a prior probability density function, and $\alpha_\chi,\beta_\chi$ are hyper-parameters\footnote{For convenience, we later re-parameterize the Beta distribution in terms of the spin magnitude mean $\mu_\chi$ and standard deviation $\sigma_\chi$.} controlling the shape of the Beta distribution, which is defined on the interval $\chi\in[0,1)$ and constrained to only allow non-singular distributions ($\alpha_\chi,\beta_\chi \geq 1$).
The spin tilt distribution, meanwhile, is from \cite{spin}:
\begin{align}\label{eq:z}
    \pi(z_{1,2} | \zeta, \sigma_t) = \zeta \,  G_t(z_{1,2} | \sigma_t) + 
    (1-\zeta) \, \mathfrak{I}(z_{1,2}) ,
\end{align}
where $z_{1,2}\equiv \cos \theta_{1,2}$ and $\theta_{1,2}$ are the misalignment angles between the orbital angular momentum vector and the respective spin vectors. Here $G_t$ is a Gaussian distribution with a peak at $z=1$ (aligned) and a width of $\sigma_t$, truncated at $z=[-1,1]$.
This sub-model represents a population of preferentially aligned field binaries.
Meanwhile, $\mathfrak{I}$ is a uniform distribution on the interval $z\in[-1,1]$.
This sub-model represents an isotropic distribution associated with dynamical formation.
The hyper-parameter $\zeta$ is a mixing fraction determining the relative importance of each sub-model.\footnote{Other possible formation channels (e.g., hierarchical triple systems, formation in active galactic nuclei, primordial black holes) have distinct spin predictions.  However, this model focuses on the isolated binary and dynamical formation channels.}
The final two spin degrees of freedom $\phi_1,\phi_2$ describing the azimuthal directions of the spin vectors are assumed to be uniformly distributed.

In this paper we do not carry out calculations using the second model from \cite{o3a_pop}---the \gaussian model \citep{Roulet2019, Miller2020}.
However, we argue below that it shares key features with the \default model.   We therefore expect that our findings obtained with the \default model are applicable to results obtained with the \gaussian model as well.

Using the \default model, \cite{o3a_pop} reconstruct the distribution for the effective inspiral spin parameter
\begin{align}
    \chi_\text{eff} \equiv 
    \frac{\chi_1\cos\theta_1 + q\chi_2\cos\theta_2}{1+q} ,
\end{align}
where $q\equiv m_2/m_1 \leq 1$ is the mass ratio, the distribution of which is modeled according to a power law:
\begin{align}\label{eq:q}
    \pi(q | \beta_q) \propto q^{\beta_q} .
\end{align}
The $\chi_\text{eff}$ parameter is frequently used to interpret results because it is an approximate constant of motion in precessing binaries and it is a relatively well-measured quantity.
\cite{o3a_pop} find that $12-44\%$ of BBH systems possess $\chi_\text{eff}<0$ (90\% credible interval), indicating that at least one black hole tilt angle $>90^\circ$.

However, this finding is disputed by \cite{Roulet2021}.
These authors analyze a slightly different set of events (chosen to be more confidently detected) using a distinct set of posterior samples.
They carry out two analyses of black hole spin.
We focus for the moment on their ``model-free'' analysis, which assumes that $\chi_\text{eff}$ is uniformly distributed on the interval $[-1,1]$.
The prior for the poorly-measured variable
\begin{align}
    \chi_\text{diff} \equiv \frac{q\chi_1\cos\theta_1 - \chi_2\cos\theta_2}{1+q} ,
\end{align}
is chosen to be uniform, conditioned on $\chi_\text{eff}$ and subject to the constraint that $\chi_{1,2} \leq 1$ as required by general relativity.\footnote{Given these constraints, the marginalised prior on $\chi_\text{diff}$ is not actually uniform.}
This leaves four degrees of freedom $\chi_{ix},\chi_{iy}$ (where $i=1,2$), which are assumed to be uniformly distributed on a disk with radius $\sqrt{1-\chi_{iz}^2}$.
We call this set of assumptions \textsc{Roulet+}.

\begin{figure}
    \centering
    \includegraphics[width=\columnwidth]{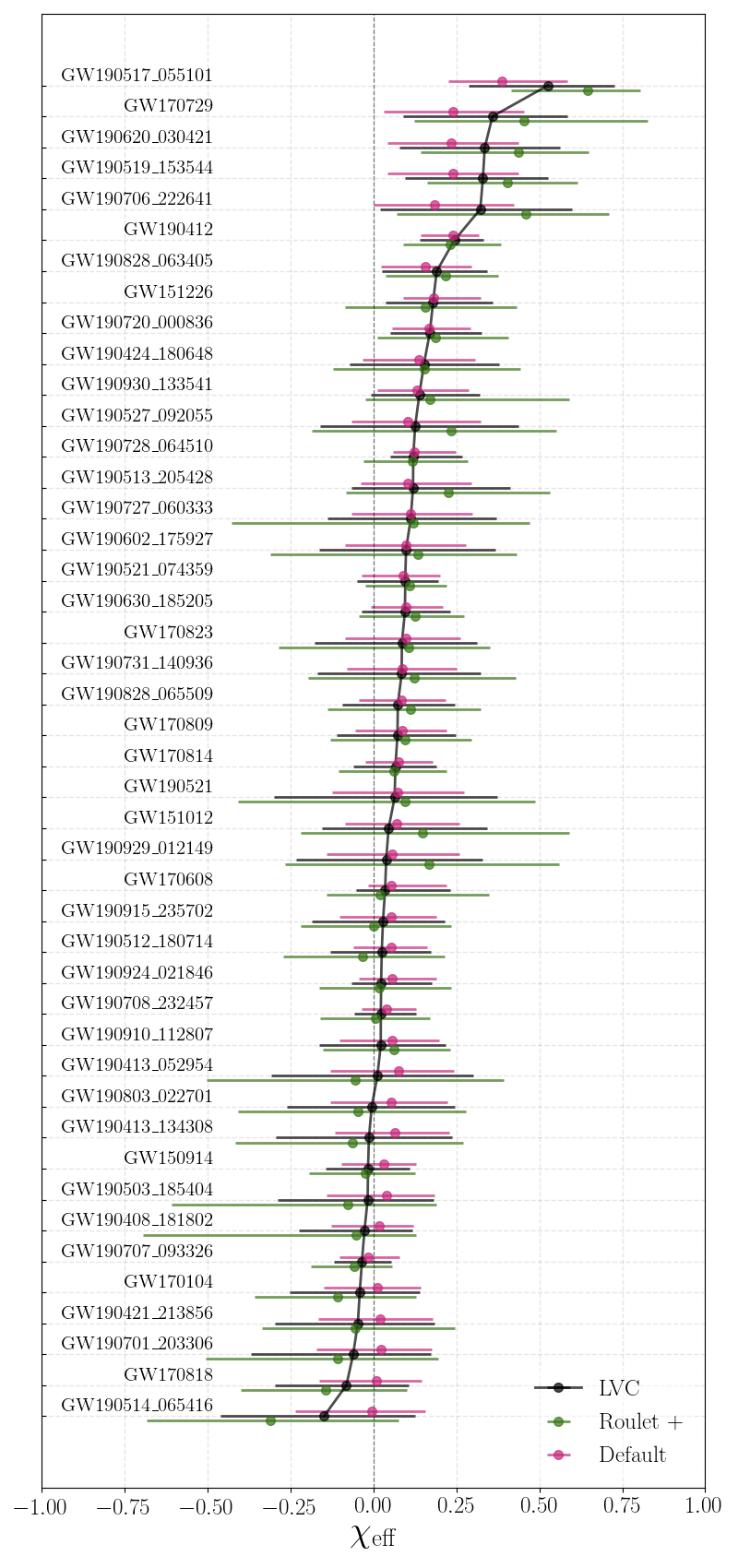}
    \caption{
    A ``dot plot'' showing the medians and 90\% credible intervals for $\chi_\mathrm{eff}$ assuming different population models.
    The fiducial model used by LIGO--Virgo (LVC, black) assumes uniform distribution for dimensionless spins $\chi_1, \chi_2$ with an isotropic distribution for the spin directions.
    Green (\textsc{Roulet+}) shows the model from \cite{Roulet2021}. The \default model from \cite{o3a_pop} is shown in pink.
    No models indicate support for a sub-population of events with $\chi_\text{eff}<0$.
    }
    \label{fig:dot-plot}
\end{figure}

Using these distributions, \cite{Roulet2021} plot the credible intervals for $\chi_\text{eff}$ for individual events in their catalog.
We reproduce this ``dot plot'' in Fig.~\ref{fig:dot-plot} for the events in  \cite{o3a_pop}, using posterior samples from that study \footnote{Posterior samples from \href{https://dcc.ligo.org/LIGO-P2000223/public}{dcc.ligo.org/LIGO-P2000223/public}}.
Individual event posteriors computed using priors from different population models are represented with different colors: black is the fiducial LIGO--Virgo spin prior (uniform in $\chi_1, \chi_2$ with an isotropic prior for the spin directions), green is the population model from \cite{Roulet2021}, and pink is the \default model from \cite{o3a_pop} (averaged over the posterior on the population parameters).

By inspecting a version of this plot, \cite{Roulet2021} find there is not clear support for a population of events with $\chi_\text{eff}<0$.
Under all three population models, there are no events which have exclusively negative $\chi_\text{eff}$ support inconsistent with zero.
We therefore concur with \cite{Roulet2021} that the BBH mergers observed to date are consistent with a sub-population of negligible-spin events $\chi_{1,2}\approx0$ with a second sub-population of events with $\chi_\text{eff}>0$.
The signature from \cite{o3a_pop} for a sub-population of $\chi_\text{eff}<0$ seen by \cite{o3a_pop} is likely due to a misspecification of the assumed population spin model.
This paper therefore seeks to improve on the \default model by developing a more sophisticated model for black hole spin in order to better describe the data and to provide more reliable inferences.

The remainder of this paper is structured as follows.
In Section~\ref{diagnosis}, we discuss why checks in \cite{o3a_pop}, designed to validate the observation of a sub-population with $\chi_\text{eff}<0$, did not reveal the model-dependency of this result.
In Section~\ref{extended}, we propose an improved black-hole spin model (called \exdefault) designed to address the shortcomings of the \default model.
We repeat the black hole spin distribution analysis from \cite{o3a_pop} (Section~\ref{sec:analysis}) using the \exdefault model and report results in Section~\ref{sec:results}.
In agreement with \cite{Roulet2021}, we show that the the observation of a sub-population with $\chi_\text{eff}<0$ is model-dependent.
We present revised fits to the distribution of black hole spins.
We conclude in Section~\ref{sec:discussion} by discussing strategies for future development and testing of black hole spin models.

\section{Diagnosing limitations of the \\
\default model}\label{diagnosis}
\cite{o3a_pop} include several checks, which would seem at first glance to support the claim for a sub-population of events with $\chi_\text{eff}<0$.
First, they note the similar reconstructed distributions for $\chi_\text{eff}$ obtained with two different models: the \default model described above and \gaussian model that describes the distribution of $\chi_\text{eff}$ and the effective precession spin parameter $\chi_p$ using a multi-variate Gaussian (see their Fig.~11b).
The potential for model-induced systematic error is reduced given the signature is visible using two different models.
Second, they carry out posterior predictive checks to show that mock source populations drawn from the \gaussian model would have similar distributions of $\chi_\text{eff}$ and $\chi_\text{p}$ to the observed data; see their Fig.~26.
Third, they perform an analysis in which a new hyper-parameter $\chi_\text{eff}^\text{min}$ is added to the population model, which enforces a minimum value of $\chi_\text{eff}>\chi_\text{eff}^\text{min}$; see their Fig.~27.
A negative value for $\chi_\text{eff}^\text{min}$ is preferred over $\chi_\text{eff}^\text{min}=0$, seemingly bolstering the case for a sub-population of events with $\chi_\text{eff}<0$.

It is instructive to discuss how each of these checks fails to catch the apparent lack of $\chi_\text{eff}<0$ events in Fig.~\ref{fig:dot-plot}.
While the agreement between the \default and \gaussian models provides something of a sanity check, both models lack a key feature: the ability to account for an excess of BBH systems with $\chi_{1,2}\approx0$.
The possibility that some fraction of LIGO--Virgo binaries should merge with negligible spin is supported by theoretical studies of angular momentum transport~\citep{Fuller2019b, Belczynski:2020}.
Moreover, the negligible spin of many/most LIGO--Virgo binaries has been noted observationally \citep{Miller2020,gwtc2_hierarchical}.
Neither spin model from \cite{o3a_pop} accounts for a sub-population of binaries with $\chi\approx0$ black holes.
Since neither model can accommodate an excess of events with $\chi_{1,2}\approx0$, the models are liable to fit such an excess with the next best thing: a subpopulation of events distributed \textit{about} $\chi_\text{eff}=0$.
In this way, the \default and \gaussian models can both yield false positive signals for $\chi_\text{eff}<0$ when the true population contains an excess of events with $\chi_{1,2}\approx 0$.

Next we turn to the posterior predictive check from \cite{o3a_pop}, their Fig.~26.
This plot compares the cumulative distribution of $\chi_\text{eff}$ for the data to the distribution expected from the model.
The two distributions are visually consistent.
While a mismatch between data and model predictions would indicate a failure of the model, a match does not prove that the model is a faithful representation of the data.
Using a different population model, \cite{Roulet2021} find no evidence for a sub-population of events with $\chi_\text{eff}<0$---a finding that we confirm independently below.

Finally, we consider the test from \cite{o3a_pop} showing that the data prefer a negative values of $\chi_\text{eff}^\text{min}$ such that some events in the population are characterized by negative $\chi_\text{eff}$ values on the interval $(\chi_\text{eff}^\text{min},0)$ (see their Fig.~27).
It should now be apparent that this test can be tricked if the data contain an excess of events with $\chi_{1,2}\approx 0$.
In order to illustrate this---and to build an improved model for BBH spin---we introduce an \exdefault model, which incorporates elements from \cite{Roulet2021} that enable an excess of negligible-spin events.

\section{The \exdefault model}\label{extended}
We introduce two changes to the \default model.
First, we add a new population parameter $\lambda_0$,
corresponding to the fraction of BBH mergers with negligible spin.
Our revised spin magnitude distribution is
\begin{align}\label{eq:chi_extended}
    \pi(\chi_{1,2} |  \alpha_\chi, \beta_\chi, \lambda_0) = & (1-\lambda_0)\text{Beta}(\chi_{1,2} | \alpha_\chi, \beta_\chi) \nonumber\\
    & + \lambda_0 G_{\mathrm{t}}\left(\chi_{1,2} \mid \mu=0, ~\sigma_0\right) ,
\end{align}
where $G_t$ is a truncated Gaussian distribution peaking at $\chi_{1,2}=0$ with width $\sigma_0$.
We set $\sigma_0=0$ so that the spin magnitudes are precisely zero and carry out the analysis using a dedicated set of zero-spin posterior samples.
Setting $\chi=0$ is probably a good approximation since \cite{Fuller2019b} suggest that typical black-hole spin magnitudes are $\chi\approx 0.01$ (ignoring binary effects, particularly tidally induced spin-up, discussed by, e.g., \citealt{Kushnir:2016,Zaldarriaga:2018,Qin2018,Bavera2020,Belczynski:2020,MandelFragos:2020}).
It would be interesting to make $\sigma_0$ a population parameter that can be fit with the data.
However, we leave this for future work as there are some technical challenges; preliminary studies with small, non-zero values of $\sigma_0$ seem to suffer from under-sampling effects.
We adopt a uniform prior for $\lambda_0$ on the interval $[0,1]$.

Our second change is to add a population parameter $z_\text{min}$ to the distribution of black-hole tilt angles such that the combined distribution from Eq.~\ref{eq:z} is forced to zero for $z_{1,2}<z_\text{min}$:
\begin{align}\label{eq:z_extended}
    \pi( z_{1,2}& |  \zeta, \sigma_t, z_\text{min}) \propto \nonumber\\
    & \Big( \zeta \,  G_t(z_{1,2} | \sigma_t) + 
    (1-\zeta) \, \mathfrak{I}(z_{1,2}) \Big) \Theta(z_{1,2}-z_\text{min}) .
\end{align}
We adopt a uniform prior for $z_\text{min}$ on the interval $[-1,1]$.
In section \ref{sec:results} we present results of an analysis of events from the GWTC-2 catalog using the \exdefault model.
The full list of population hyper-parameters and priors is given in Table~\ref{tab:parameters_default}.

\begin{table*}
    \centering
    \begin{tabular}{lll} 
        \hline
        {\bf Parameter} & \textbf{Description} & \textbf{Prior} \\\hline\hline
        $\lambda_0$ & Mixing fraction of mergers with negligible spin, $\chi_{1,2}\lesssim\sigma_0$  & U(0,1) \\
        $\sigma_0$ & Spread in $\chi_{1,2}$ for systems with negligible spin & $\sigma_0=0$ \\
        $\mu_\chi$ & Mean of spin magnitude distribution &  U(0,1) \\
        $\sigma_\chi^2$ & The square of the width of the spin magnitude distribution  & U(0,0.25) \\
        $\zeta$ & Mixing fraction of mergers with preferentially aligned spin &  U(0,1)\\
        $\sigma_t$ &  Spread in projected misalignment for preferentially aligned black holes & U(0,4) \\
        $z_\text{min}$ & Minimum value of the projected misalignment & U(-1,1) \\
        \hline
    \end{tabular}
    \caption{
    Summary of \exdefault model hyper-parameters.
    The notation $U(a,b)$ indicates a uniform distribution on the interval ranging from $a$ to $b$.
    }
  \label{tab:parameters_default}
\end{table*}

\section{Analysis}\label{sec:analysis}
We use \texttt{GWPopulation} \citep{gw_population} to obtain posterior distributions for the population parameters in the \exdefault model using the same event list used in \cite{o3a_pop}.
This dataset consists of 44 confidently detected BBH mergers.
It does not include GW190814 \citep{GW190814}, which may be a neutron-star black-hole binary and which is a clear outlier from the rest of the population.
The analysis from \citet{o3a_pop} employed higher-order modes waveforms, which we are unable to use here for technical reasons. Instead, we use \textsc{IMRPhenomPv2} and \textsc{IMRPhenomD} waveforms \citep{Hannam2014, Husa2016, Khan2016} to obtain samples for $\chi>0$ and $\chi=0$ respectively. While this choice of waveforms is unlikely to affect our main conclusions, the different choice of waveform leads to subtle shifts in the posterior distribution of some population parameters. We estimate these shifts by comparing the \default model results with \textsc{IMRPhenomPv2} waveforms to the \default model results with higher-order mode waveforms. We determine that the typical values of $\chi$ for the sub-population with spinning black holes would likely lower by $\sim 0.05$ if we had used higher-order mode waveforms. Additional details are available in the companion repository.\footnote{Supplementary material including analysis inputs, posterior samples and additional plots are available here: \href{https://github.com/shanikagalaudage/bbh_spin}{https://github.com/shanikagalaudage/bbh\_spin}}
The \texttt{GWPopulation} package employs \texttt{Bilby} \citep{bilby,bilby_gwtc1} and \texttt{dynesty} \citep{dynesty}.
We fit the distribution of black-hole masses using the \textsc{Power Law + Peak} model from \cite{o3a_pop} adapted from \cite{mass}.
We fit the merger redshift distribution using the \textsc{Power-Law Evolution} model from \cite{o3a_pop}, adapted from \cite{Fishbach2018}.
We employ two sets of posterior samples: one using the standard LIGO--Virgo priors and one generated using a zero-spin prior; see \citet{gwtc2_hierarchical}.\footnote{Astute readers may notice that \cite{gwtc2_hierarchical} reports that the fraction of BBH systems with negligible spin is consistent with zero $\lambda_0\approx0$. In that study, the BBH events with measurable spin are likely attributed to a sub-population containing one or more rapidly spinning ``second-generation'' black holes (formed from previous mergers). Thus, they likely do not influence the fit for first-generation dimensionless spin parameters, $\alpha_\chi, \beta_\chi$, which are found to be consistent with a population of low-spin black holes.}
The zero-spin samples are necessary to avoid artifacts due to under-sampling.
Additional technical details are provided in Appendix~\ref{details}.

We adopt the same treatment of selection effects as used in \cite{o3a_pop}, which accounts for mass-dependent Malmquist bias, but not selection effects due to spin.
The authors of that work explain that they cannot reliably estimate spin-induced selection effects using the currently available injection set; see their Appendix~F.
It is slightly easier to detect BBH signals with $\chi_\text{eff}>0$, so we expect that our fit may slightly over-emphasize high positive values of $\chi_\text{eff}$ relative to the true distribution.
Fig. 1 of \cite{Ng2018} illustrates this shift in the inferred  $\chi_\mathrm{eff}$ distribution. This shift will impact the branching ratio between the two non-spinning and spinning sub-populations. We estimate that including selection effects would decrease the estimated sub-population of spinning BBHs by $\lesssim 20\%$ of the currently estimated fraction $(1-\lambda_0)$, i.e., $\lesssim 4\%$. Thus, the systematic error from selection effects is less than the current statistical uncertainty.

\section{Results}\label{sec:results}
\begin{figure*}
    \includegraphics[width = 0.49\textwidth]{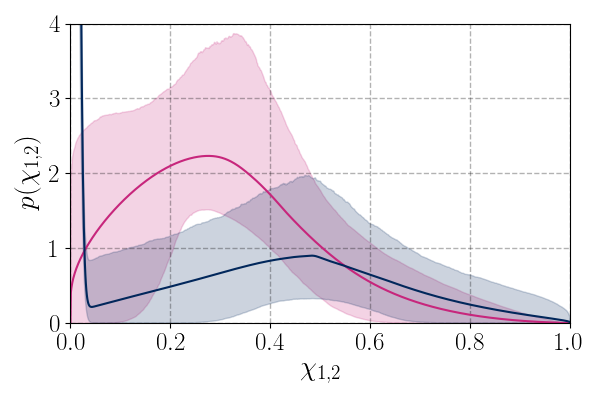} 
    \hspace{1mm}
    \includegraphics[width = 0.49\textwidth]{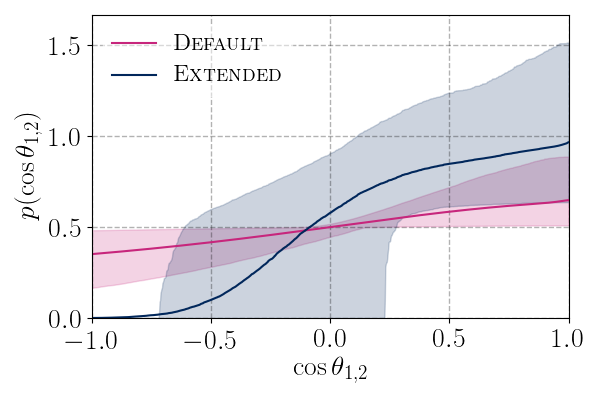}
    \caption{
    Population predictive distributions for the \default (pink) and the \exdefault (navy) models.
    \textit{Left:} The distribution of dimensionless spin. \textit{Right:} The distribution of the cosine of the tilt angle.
    The solid curves represent the mean and the shaded region represents the 90\% credible interval. 
    In the left-hand panel, we represent the \exdefault model's delta function at $\chi=0$ with a narrow Gaussian with a width of $\sigma_0=0.01$ for visibility and is zoomed in.
    }
    \label{fig:default_compare_ppds}
\end{figure*}

We plot population predictive distributions (PPDs)\footnote{The PPD is given by the conditional prior marginalized over the posterior distribution of the population parameter
\begin{align}
    p_\Lambda(\chi_{1,2} | d) = \int d\Lambda \, 
    p(\Lambda | d) \,
    \pi(\chi_{1,2} | \Lambda) .
\end{align}
Here, $\Lambda$ are population parameters described in Table~\ref{tab:parameters_default}.}
comparing the \default and \exdefault models in Fig.~\ref{fig:default_compare_ppds}.
In the left panel, we show the reconstructed spin magnitude distribution, the \exdefault model exhibits clear support for a narrow peak at $\chi_{1,2}=0$, which is not present in the \default model because the model lacks the flexibility to fit this peak. 
In the right panel, we show the reconstructed distribution of the (cosine of the) spin tilts, we see that the \exdefault model distribution of $z_{1,2}\equiv\cos\theta_{1,2}$ tapers off for $\cos\theta_{1,2}\lesssim 0$ while the \default model exhibits considerably more support for binaries with $\cos\theta_{1,2}<0$. 
Since the \exdefault model includes the \default model as a special case (with $\lambda_0=0, z_\text{min}=-1$), we conclude that the data prefer a sub-population of binaries with $\chi_{1,2}\approx0$ over a sub-population of binaries with $\chi_\text{eff}<0$.
This conclusion is supported by our model selection results, summarized in Table~\ref{tab:bf}, which show the \exdefault model is preferred over the \default model by {$\log_{10}{\cal B}\approx{3.55}$}.  This preference clearly comes from the introduction of a zero spin sub-population (cf.~the $\lambda_0=0$ model). There is a slight preference for the \exdefault model with a broad uniform prior on $z_\text{min}$ or with $z_\text{min}$ fixed to zero than with $z_\text{min}$ fixed to $-1$.

\begin{figure*}
    \includegraphics[width = 0.49\textwidth]{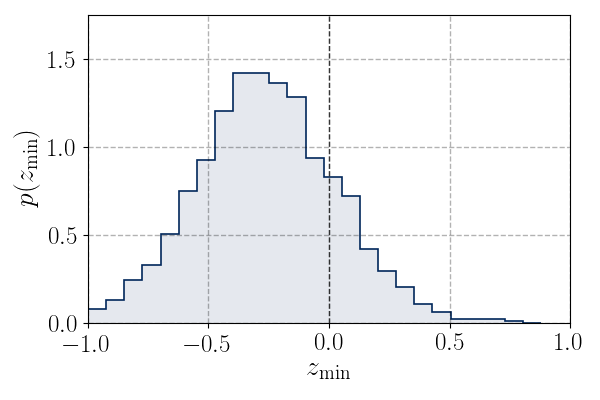} 
    \hspace{1mm}
    \includegraphics[width = 0.49\textwidth]{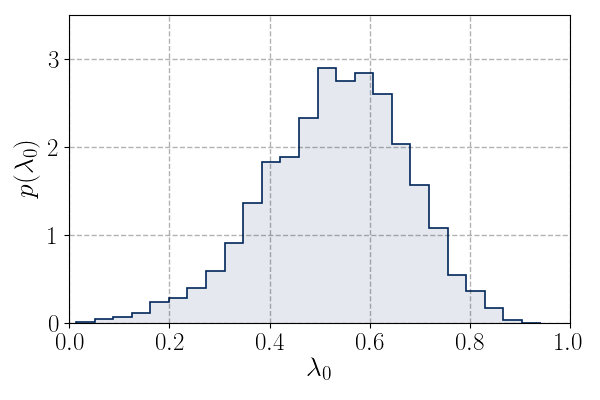}
    \caption{
    Posterior distributions for population parameters new to the \exdefault model.
    \textit{Left:} The posterior distributions for the population parameter $z_\mathrm{min}$, which determines the minimum cosine tilt angle such that $\cos\theta_{1,2}\geq z_\text{min}$.
    \textit{Right:} The posterior distribution for $\lambda_0$, the fraction of binaries with negligible black-hole spins $\chi_{1,2} = 0$.
    }
    \label{fig:z_min}
\end{figure*}

Figure~\ref{fig:z_min} shows the posterior distributions for the two variables new to the \exdefault model: $z_\text{min}$ (left) and $\lambda_0$ (right).
Turning first to the posterior for $z_\text{min}$, we find ample posterior support for $z_\text{min}=0$, which means the data are consistent with the premise that all merging binaries have $\chi_\text{eff}>0$.
Next we turn our attention to the posterior distribution for $\lambda_0$ (right), the parameter that controls the fraction of binaries merging with negligible black-hole spin.
The distribution shows that {majority of} BBH systems merge with negligible spins: {$\lambda_0 = 0.54_{-0.25}^{+0.21}$} (90\% credibility).

\begin{table}
\begin{tabular}{l c c}
    \hline
    Spin model & $ \log_\mathrm{10}{\cal B}$ & $\Delta \log_\mathrm{10}{\cal L}_\mathrm{max}$ \\
    \hline\hline
    \default & 0.00 & 0.00 \\
    \exdefault & {3.55} & {3.68}\\ \hline
    \exdefault with $z_\text{min}=0$ & {3.94} & {3.33} \\
    \exdefault with $z_\text{min}=-1$ & {2.87} & {3.03}\\
    \exdefault with $\lambda_0=0$ & {1.09} & {0.80}\\
    \hline
\end{tabular}
\caption{
Log Bayes factors and maximum log likelihood differences for different spin models compared to the \default model.
}
\label{tab:bf}
\end{table}

\begin{figure*}
    \centering
    \includegraphics[width=1.5\columnwidth]{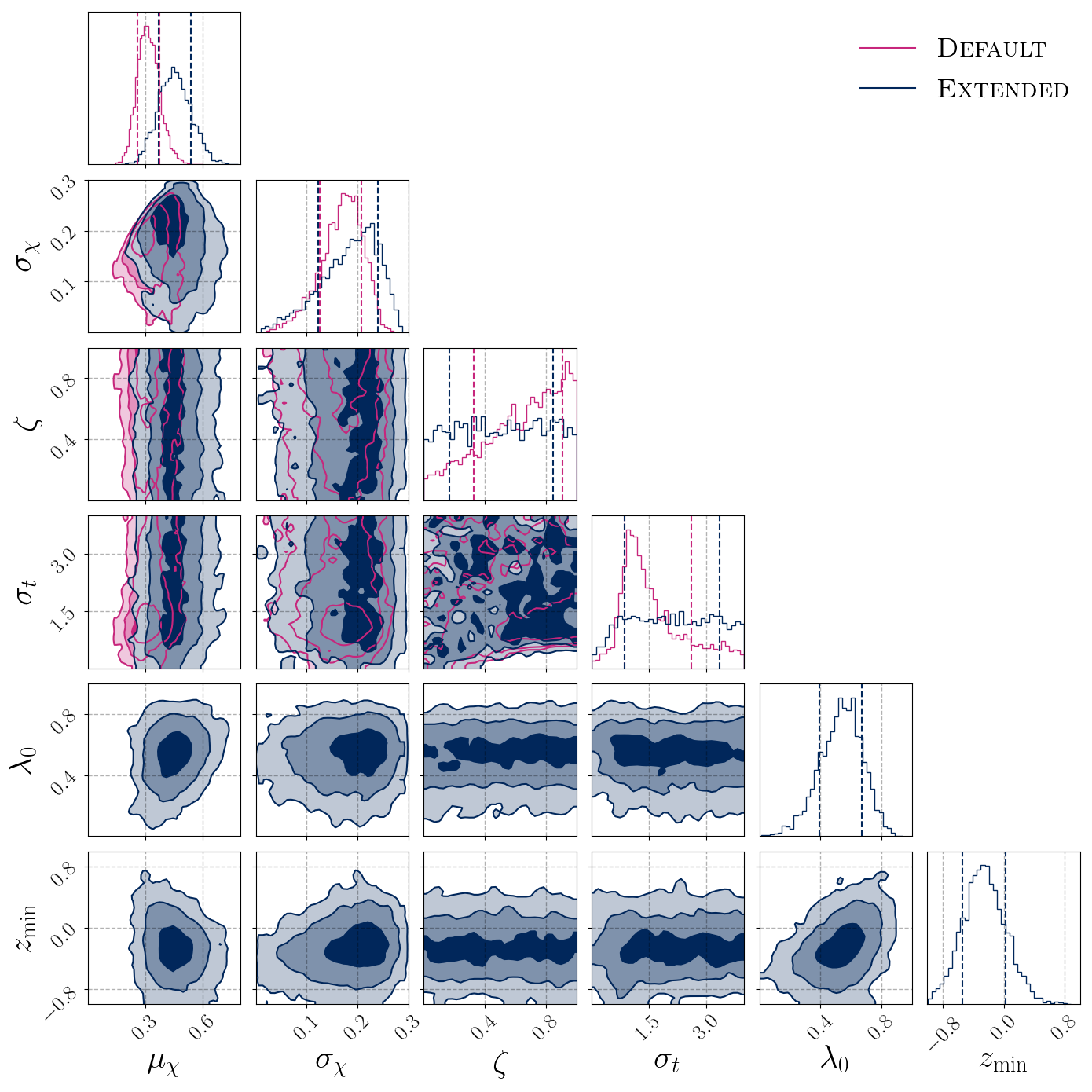}
    \caption{
    A corner plot showing hyper-posterior distributions for the \default (pink) and \exdefault (navy) models.
    The parameters are summarized in Table~\ref{tab:parameters_default}.
    The shaded regions represent one-, two-, and three-sigma credible intervals.
    }
    \label{fig:default_compare_posteriors}
\end{figure*}

In Fig.~\ref{fig:default_compare_posteriors} we include a corner plot showing the posterior distribution for all the population parameters in the \exdefault model. 
The navy contours mark the one-, two-, and three-sigma credible intervals for the \exdefault model while the pink contours show the results for the \default model used in \cite{o3a_pop}.
There are a number of interesting differences between the two models.
First, the $\mu_\chi$ parameter, which determines the average dimensionless spin magnitude for black holes (those with non-negligible spin in the \exdefault model), shifts from $\approx0.3$ for \default to {$\approx0.45$} for \exdefault.
This shift reflects the fact that dimensionless spin is higher when we allow for a separate population of binaries with negligible black-hole spin.
{Taken together, Figs.~\ref{fig:z_min} and \ref{fig:default_compare_posteriors} suggest two sub-populations: one population with BBH systems merging with negligible spins and the other merging with moderately large spins.}

Second, the posterior for the $\zeta$ parameter (shown in Fig.~\ref{fig:default_compare_posteriors}), which determines the fraction of binaries with preferentially aligned spin, becomes broader, approaching a uniform distribution.
This change is explained by the introduction of the $z_\text{min}$ parameter, which provides a new means of building a population of preferentially aligned binaries.\footnote{Consider, for example, the case where $z_\text{min}\approx0$ and $\zeta=0$, which is well-supported by the data. This case corresponds to a population where black hole tilts are ``half-isotropically'' distributed for angles $\theta_{1,2}<90^\circ$, but black holes never merge with tilt angles $>90^\circ$.}
Both models disfavor perfect alignment of the black-hole spin vectors with the orbital angular momentum.
The \default model shows support for a preferentially aligned population ($\zeta>0$) but with non-zero misalignment angle spread ($\sigma_t>0$).
{Meanwhile, the \exdefault model confidently excludes  perfect alignment with $z_\text{min} < 0.43$ at 99\% credibility.}


\begin{figure}
    \centering
    \includegraphics[width=\columnwidth]{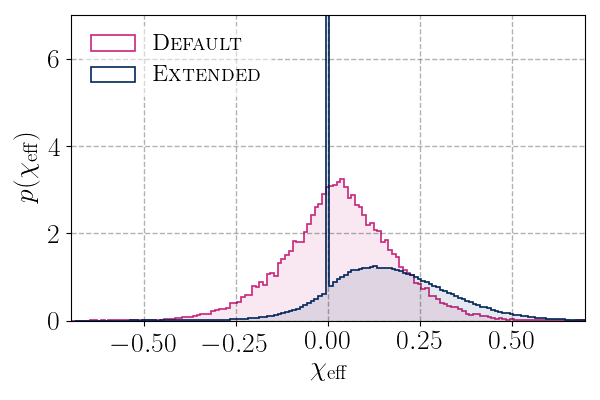}
    \caption{
    The population predictive distribution for the effective inspiral spin parameter $\chi_\mathrm{eff}$ for the \default (pink) and the \exdefault (navy) models.
    }
    \label{fig:chi_eff_default}
\end{figure}

Finally, in Fig.~\ref{fig:chi_eff_default}, we show the $\chi_\mathrm{eff}$ PPDs for the \default and \exdefault models.
The \exdefault model is characterized by a sharp peak at $\chi_\text{eff}=0$ corresponding to the sub-population of BBH systems with negligible spin.
While the \exdefault PPD does not vanish for $\chi_\text{eff}<0$, there is very limited support there, with fewer than {2\%} of all binaries predicted to have $\chi_\text{eff}<-0.1$.
This is consistent with Fig.~\ref{fig:dot-plot}, which shows no suggestion of events with negative $\chi_\text{eff}$.
This conclusion is also consistent with Fig.~\ref{fig:z_min}, which shows support for $z_\text{min}$ as low as {$z_\text{min}\approx -1$}.
However, the same plot shows that the data are consistent with $z_\text{min}=0$.
In other words, we cannot rule out the possibility of a small sub-population with $\chi_\text{eff}<0$, but the \exdefault model provides no evidence that such a sub-population exists. We also performed a set of population predictive checks, which shows the model is a good fit to the data; these figures are available in the companion repository. However, we emphasize that a match of the predicted and observed distributions does not guarantee the model accurately represents the data.

\section{Discussion}\label{sec:discussion}
Inspired by \cite{Roulet2021}, we reanalyze the population of merging binary black holes from \cite{o3a_pop} using a revised model for the distribution of black hole spins designed to allow for a sub-population with negligible spin.
Using the same events and the same posterior samples as \cite{o3a_pop}, we obtain results qualitatively similar to \cite{Roulet2021}, suggesting that \cite{o3a_pop}'s finding of a sub-population of binaries merging with $\chi_\text{eff}<0$ is model-dependent.
Using our new \exdefault model, we find that the data can be explained by the hypothesis that all binaries merge with black hole spin preferentially aligned with the orbital angular momentum \textit{or} with negligible spin.
These findings are consistent with GWTC-1 studies \citep{Farr2017, Farr2018, o1o2_pop, Juan2021} inferring the population of what was then 10 BBH detections is consistent with mostly negligible spins and a few events with support for $\chi_\mathrm{eff}>0$.

This result somewhat diminishes the case for dynamical mergers as a major channel for merging binaries.
However, the dynamical scenario remains a plausible explanation for binary mergers with black holes in the pair instability mass gap~\citep{GW190521}.
If GW190521 is a hierarchical merger (assembled from the products of previous mergers), perhaps black holes participating in the first generation of dynamical mergers are among the {$\approx 54\%$} of with negligible spin.
Indeed, studies of dynamical mergers in dense clusters find that small spins are required for first-generation black holes in order for their remnants to be retained to merge again \citep{hierarchical,gwtc2_hierarchical}.
If a large fraction of merging binaries are assembled dynamically, our results suggest that most dynamically merging black holes  have negligible spin.

A sub-population of binaries with negligible black-hole spins can also be accommodated within the field formation framework; see, e.g., \citet{Belczynski:2020,Bavera2020}.
If the stellar progenitors of BBH systems efficiently shed angular momentum through mass loss, it may be common for most field BBH systems to form with negligible spin.
A minority of progenitors in sufficiently tight binaries could, however, tidally spin up the stellar core of the secondary star prior to its collapse into a black hole, producing a BBH system with at least one rapidly spinning BH.
In this framework, black holes with non-negligible spins could be those whose naked helium star progenitors were tidally spun up in tight binaries. 
This possibility can be explored in the future by relaxing the assumption that the spins and spin-orbit misalignment angles of both binary components are independently drawn from the same distributions.  

Binary black hole detections in which the more massive black hole has clearly non-zero spin $\chi_1>0$ could be challenging to accommodate in the standard field framework as described above \citep{MandelFragos:2020,Qin2021}.
Several events in GWTC-2 would seem to fall into this category, including GW151226 \citep[][but see \citealt{MateuLucena:2021}]{GW151226, Chia:2021}, GW190412 \citep[][but see \citealt{MandelFragos:2020}]{GW190412,Zevin2020}, and GW190403\_051519 \citep[][though the high mass  already makes field formation unlikely for this system]{GWTC-2.1, Qin2021}.
If these systems formed in the field, it is possible that the more massive black hole formed from what was initially the secondary (lower mass) star that subsequently gained mass through accretion.
In this scenario, the secondary star could still be tidally spun up before collapsing into the more massive black hole, a scenario explored in \cite{Bavera2021, OlejakBelczynski:2021}.
Furthermore, the progenitor of the first-formed black hole can also be tidally spun up if the binary is initially very compact, as appears to be the case with high-mass X-ray binaries such as M33 X-7 \citep{Valsecchi:2010} and Cygnus X-1 \citep{Qin:2019}, though it is unclear if such systems lead to merging BBHs \citep{Neijssel:2021}.

In this work we focussed only on the black-hole spin distribution, assuming that the black-hole mass distribution is independent. However, there may be possible correlations between mass and spin, as explored by \citet{Safarzadeh2020, Tagawa2021, Callister:2021}. Examples include predictions for negative correlation between mass and spin for isolated binary evolution  \citep[e.g.,][]{Bavera2020} and mergers in young clusters \citep[e.g.,][]{Kumamoto2021}. Meanwhile, positive correlations are predicted for repeated dynamical mergers in globular clusters \citep[e.g.,][]{Rodriguez2018} with a possible high-mass spin-aligned contribution from chemically homogeneously evolving binaries \citep[e.g.,][]{Marchant2016, MandelDeMink2016, Riley:2020}, while mergers in AGN disks may exhibit a positive correlation between mass and the dispersion of effective spin \citep{Tagawa2020}. Investigating such correlations, along with possible correlations with redshift, can improve the accuracy of the models and enable them to better distinguish between various evolutionary scenarios.

Our work highlights the subtleties in diagnosing model misspecification, which can lead to overly model-dependent conclusions.
While \cite{o3a_pop} perform a number of checks to validate their evidence for a sub-population of events with $\chi_\text{eff}<0$ (see Section~\ref{diagnosis}), none of these checks pinpointed how the model could misbehave when applied to a distribution with a sub-population of binaries with negligible-spin black holes, as pointed out by \cite{Roulet2021}.
We recommend careful consideration of possible \textit{sharp features}---in this case, a narrow peak with $\chi_{1,2}\approx0$---since they can yield misleading results when fit with slowly varying functions.

The ``dot plot'' featured in Fig.~1 of \cite{Roulet2021}, and included in our own Fig. \ref{fig:dot-plot}, provides an important visual check.
However, it is not obvious that such a plot can be applied more broadly to help spot model misspecification.
For one thing, the $\chi_\text{eff}$ dot plot is fairly straightforward to interpret because the likelihood function for individual events is approximately Gaussian in this parameter.
However, this is not true for other variables, for example, $\chi_p$.
Moreover, even the $\chi_\text{eff}$ dot plot includes a hidden population model, which can affect its interpretation.

\begin{figure}
    \centering
    \includegraphics[width=\columnwidth]{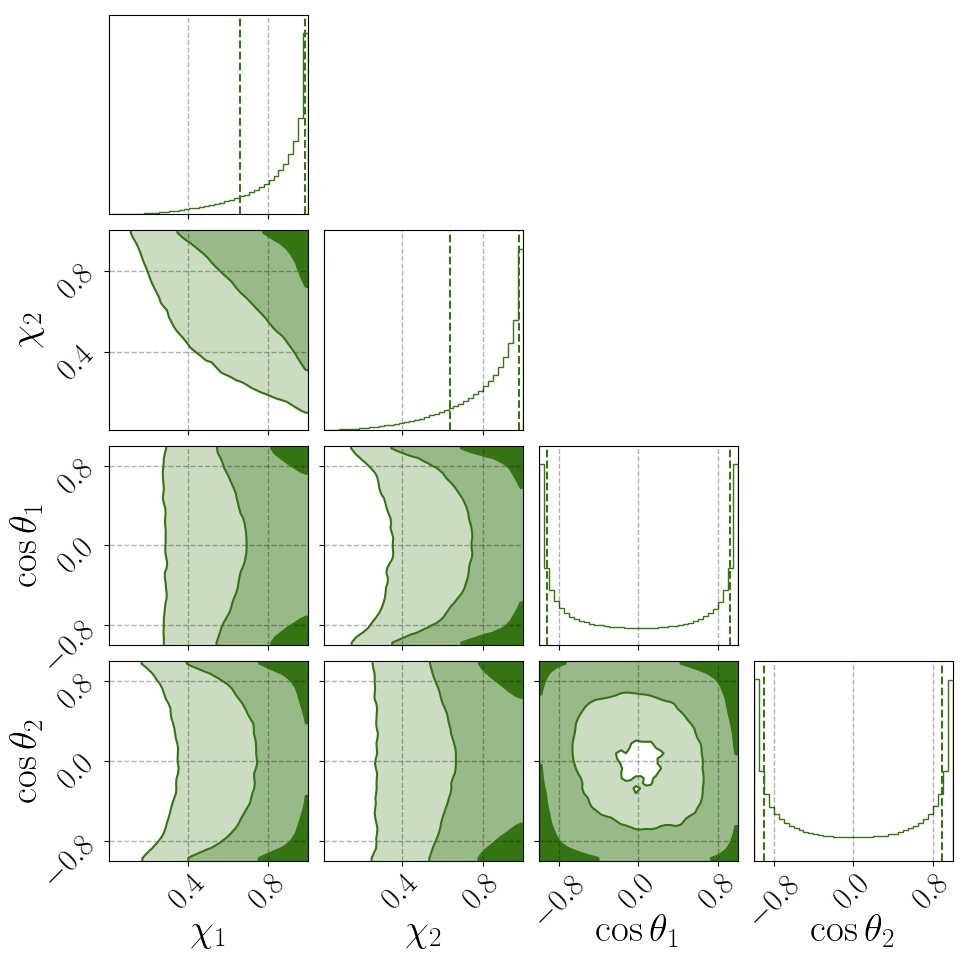}
    \caption{
    Prior distribution on spin magnitudes and cosine of the tilt angle for the \cite{Roulet2021}  $\chi_\mathrm{eff}$ model.
    }
    \label{fig:ias_priors}
\end{figure}

In order to obtain a prior distribution that is uniform in $\chi_\text{eff}$, the ``model-free'' prescription in \cite{Roulet2021} makes implicit assumptions about the distribution of the physical spin parameters $\chi_1$, $\chi_2$, $\cos \theta_1$ and $\cos \theta_2$.
Plotting the distribution of these spin parameters in Fig.~\ref{fig:ias_priors}, we see that some of the assumed distributions are not especially physical.
In particular, the ``model-free'' analysis implicitly assumes that black holes preferentially spin near the Kerr limit $\chi_{1,2}=1$ with a strong covariance between $\chi_1$ and $\chi_2$.
The distribution of spin tilts, meanwhile, favours spin vectors that are either aligned or anti-aligned with the orbital angular momentum vector.
Comparing the green ``Roulet +'' model to the black LVC model in Fig.~\ref{fig:dot-plot}, we see that the two models produce similar dot plots when one accounts for the implicit preference for larger spins in the ``model-free'' approach.
However, the two models do not produce the same ordered list when we rank events from smallest to largest $\chi_\text{eff}$, which indicates that other spin degrees of freedom described by the implicit model influence this plot -- an effect that will become more significant as the gravitational-wave catalog grows.  However, the surprising shape of the distributions in Fig.~\ref{fig:ias_priors} reminds us that all fully specified models of black-hole spin require \textit{some} distribution of physical parameters even if these distributions are not explicit.  It therefore seems useful to cast population models in terms of physical parameters, or, at the very least, to check the suitability of the distribution of physical parameters.

There are a number of interesting extensions to this work worthy of future exploration.
Our \exdefault model is designed to investigate questions raised in \cite{o3a_pop} and \cite{Roulet2021}.  A number of possible model extensions consider more astrophysically motivated distributions.  As discussed above, spin magnitudes and directions may be coupled with each other, or with mass and mass ratio (see, e.g., \citealt{Callister:2021}). For example, there could be different spin magnitude distributions in the field scenario (preferentially aligned spins) and the dynamical scenario (isotropic spins).  The low-spin-magnitude subpopulation could incorporate small but non-zero spins by varying $\sigma_0$. 
The $z_\text{min}$ parameter could be applied only to the sub-population of binaries with $\cos\theta_{1,2}$ described by the truncated Gaussian distribution; this would restore the mixture model choice between a preferentially aligned distribution and an isotropic distribution.

Note: The results in this paper differ from the published version following some fixes to our analysis. A bug in our implementation resulted in the non-spinning posteriors to be given more weight where the fiducial prior per event was 4 times larger than it should have been. The figures and tables in this paper have been updated following this fix.

\section*{Acknowledgements}

We thank Tom Callister for his helpful comments during the preparation of this manuscript. Several authors are supported by the Australian Research Council (ARC) Centre of Excellence for Gravitational-wave Discovery (OzGrav), project number CE170100004.
IM is the recipient of the ARC Future Fellowship FT190100574.
The authors are grateful for computational resources provided by the LIGO Laboratory and supported by National Science Foundation Grants PHY-0757058 and PHY-0823459. This is document LIGO-P2100318.  \software{\texttt{GWPopulation} \cite{gw_population}, \texttt{Bilby} \cite{bilby, bilby_gwtc1}, \texttt{dynesty} \cite{dynesty}}. 

\bibliography{refs}

\begin{appendix}
\section{Implementing the \exdefault model with zero-spin samples}\label{details}
In this appendix we describe our implementation of inference on the \exdefault model hyper-parameters using two sets of posterior samples---one allowing for black hole spins (called ``fiducial'' samples), the other assuming zero black hole spins.
Our starting point is the likelihood for the data $d$ associated with a single gravitational-wave event given the population hyper-parameters $\Lambda$:
\begin{align}\label{eq:hyper_likelihood}
    {\cal L}(d|\Lambda) = & \int d\chi_{1,2} 
    \int dz_{1,2} \int d\eta \,
    {\cal L}(d|\chi_{1,2}, z_{1,2}, \eta) \, 
    \pi(\chi_{1,2}| \alpha_\chi, \beta_\chi, \lambda_0) \, 
    \pi(z_{1,2} | \zeta, \sigma_t, z_\text{min}) \,
    \pi(\eta | \Lambda) .
\end{align}
Here, ${\cal L} (d|\chi_{1,2},z_{1,2}, \eta)$ is the usual Gaussian likelihood of the data conditional on spin parameters $\chi_{1,2}, z_{1,2}$ and other parameters (mass, redshift, etc.), which we denote by $\eta$.
The next term $\pi(\chi_{1,2}| \alpha_\chi, \beta_\chi, \lambda_0)$ is the \exdefault model for spin magnitude described in Eq.~\ref{eq:chi_extended}.
This is followed by $\pi(z_{1,2} | \zeta, \sigma_t, z_\text{min})$, which is the \exdefault model for spin tilts; see Eq.~\ref{eq:z_extended}.
The final term $\pi(\eta | \Lambda)$ describes the population model for mass and redshift as well as the usual priors for extrinsic parameters.
Our mass model is the \textsc{Power Law + Peak} model from \cite{o3a_pop} adopted from \cite{mass}.
Our redshift model is the \textsc{Power-Law Evolution} model from \cite{o3a_pop} adopted from \cite{Fishbach2018}.

The likelihood can be written in terms of a non-zero-spin likelihood and a zero-spin likelihood:
\begin{align}
    {\cal L}(d|\Lambda) = 
    (1-\lambda_0) {\cal L}_{\chi>0}(d|\Lambda)
    + 
    \lambda_0 {\cal L}_{\chi=0}(d|\Lambda)
\end{align}
where
\begin{align}
    {\cal L}_{\chi>0}(d|\Lambda) = & \int d\chi_{1,2} 
    \int dz_{1,2} \int d\eta \,
    {\cal L}(d|\chi_{1,2}, z_{1,2}, \eta) \, 
    \pi(\chi_{1,2}| \alpha_\chi, \beta_\chi) \, 
    \pi(z_{1,2} | \zeta, \sigma_t, z_\text{min}) \,
    \pi(\eta | \Lambda) \\
    {\cal L}_{\chi=0}(d|\Lambda) = & \int d\chi_{1,2} 
    \int dz_{1,2} \int d\eta \,
    {\cal L}(d|\chi_{1,2}, z_{1,2}, \eta) \, 
    \delta(\chi_{1,2}) \, 
    \pi(z_{1,2} | \zeta, \sigma_t, z_\text{min}) \,
    \pi(\eta | \Lambda) \\
    = & \int d\eta \,
    {\cal L}(d|\chi_{1,2}=0, z_{1,2}, \eta) \, 
    \pi(\eta | \Lambda) . 
\end{align}
Using the standard trick for ``recycling'' posterior samples; see, e.g., \citep{intro}. In order to recycle, we rewrite the likelihood in terms of the posterior $p$ the fiducial evidence $\pi(\theta|\O)$, and the fiducial evidence ${\cal Z}_{\O}$ so that
\begin{align}
    {\cal L}_{\chi>0}(d|\Lambda) = 
    \frac{p_{\chi>0}(\Lambda|d)}{\pi(\theta|\O)}
    {\cal Z}_{\O} .
\end{align}
Then we use the fact that
\begin{align}
    \int d\theta \, p(\theta) f(\theta) = 
    \frac{1}{n_s} \sum_k f(\theta_k) ,
\end{align}
to rewrite the integral as a sum over posterior samples.
We rewrite ${\cal L}_{\chi>0}(d|\Lambda)$ as a sum over $n_s$ fiducial ($\chi>0$) posterior samples:
\begin{align}\label{eq:reweighted}
    {\cal L}_{\chi>0}(d|\Lambda) = & 
    \frac{1}{n_s} \sum_k \left(\frac{\text{Beta}(\chi_{1,2}^k|\alpha_\chi, \beta_\chi)}{\pi(\chi_{1,2}^k|\O)}\right) 
    \left(\frac{\pi(z_{1,2}^k|\zeta,\sigma_t,z_\text{min})}{\pi(z_{1,2}^k|\O)} \right)
     \left(\frac{\pi(\eta^k, ... | \Lambda)}{\pi(\eta^k, ...|\O)} \right) 
     {\cal Z}_{\O} .
\end{align}
Here, ${\cal Z}_{\O}$ is the fiducial evidence obtained using the fiducial ($\chi>0$) prior.
Likewise, the zero-spin likelihood can be written in terms of the $n_0$ zero-spin samples:
\begin{align}
    {\cal L}_{\chi=0}(d|\Lambda) = & 
    \sum_j
    \left(\frac{\pi(\eta^j, ... | \Lambda)}{\pi(\eta^j, ...|\O)}\right)
    {\cal Z}_0 .
\end{align}
Here, ${\cal Z}_0$ is the zero-spin evidence obtained using the zero-spin ($\chi=0$) prior.
Putting everything together, we obtain
\begin{align}
    {\cal L}(d|\Lambda) = & 
    \frac{1-\lambda_0}{n_s} \sum_k \left(\frac{\text{Beta}(\chi_{1,2}^k|\alpha_\chi, \beta_\chi)}{\pi(\chi_{1,2}^k|\O)}\right) 
    \left(\frac{\pi(z_{1,2}^k|\zeta,\sigma_t,z_\text{min})}{\pi(z_{1,2}^k|\O)} \right)
     \left(\frac{\pi(\eta^k, ... | \Lambda)}{\pi(\eta^k, ...|\O)} \right) 
     {\cal Z}_{\O} 
     + \frac{\lambda_0}{n_0}
    \sum_j
    \left(\frac{\pi(\eta^j, ... | \Lambda)}{\pi(\eta^j, ...|\O)}\right)
    {\cal Z}_0 \nonumber\\
    = & (1-\lambda_0) \overline{w_{\O}}(\Lambda)
     {\cal Z}_{\O} 
     + \lambda_0 \overline{w_0}(\Lambda)
    {\cal Z}_0 \nonumber\\
    = & (1-\lambda_0) 
     {\cal Z}_{\O}'(\Lambda) 
     + \lambda_0 {\cal Z}_0'(\Lambda) .
\end{align}
Here, $\overline{w_{\O}}(\Lambda), \overline{w_0}(\Lambda)$ are the average importance-sampling weights for the spinning and non-spinning models respectively.
Meanwhile, ${\cal Z}_{\O}'(\Lambda), {\cal Z}_0'(\Lambda)$ are the population-weighted Bayesian evidence values for the two sub-populations.

\end{appendix}

\end{document}